\documentclass{aastex}          
\usepackage{spr-astr-addons}    
\usepackage{url}

\RequirePackage{color}

\usepackage{graphicx}
\usepackage[colorlinks=true,citecolor=blue,linkcolor=blue]{hyperref}

\begin{document}

\title{The most compact bright radio-loud AGN -- I. A new target sample selected for the space VLBI}
\shorttitle{The most compact radio AGN I.}
\shortauthor{An et al.}

\author{Tao~An\altaffilmark{1,2} } \affil{Shanghai Astronomical Observatory, Chinese Academy of Sciences, 80 Nandan Road, 200030 Shanghai, P.R. China, antao@shao.ac.cn}
\and 
\author{Xiaocong~Wu\altaffilmark{1,2} } \affil{Key Laboratory of Radio Astronomy, Chinese Academy of Sciences, China}
\and
\author{S\'andor~Frey\altaffilmark{3} } \affil{F\"OMI Satellite Geodetic Observatory, P.O. Box 585, H-1592 Budapest, Hungary}

\begin{abstract}
We investigated the archival ground-based VLBI images of the extragalactic radio sources included in both the {\it Wilkinson Microwave Anisotropy Probe (WMAP)} and the {\it Planck} catalogues, and selected 49 bright and compact sources as potential targets for space Very Long Baseline Interferometry (VLBI) observations at mm wavelengths. These sources have a flat radio continuum spectrum between 33 and 94~GHz. They are identified as core-dominated active galactic nuclei (AGN), located at declinations above $-40\degr$, and have never been observed with ground-based VLBI at 86~GHz. The radio properties of the 49 new sources are presented. We compare this new sample with similar samples of compact AGN available from earlier studies. The new candidates, together with the existing bright compact AGN sample identified from 86-GHz ground-based VLBI imaging surveys, form a catalogue of more than 160 AGN. These could be primary targets for mm-VLBI observations on the ground, as well as for future mm-wavelength space VLBI missions such as the project with two satellites currently under study in China.
\end{abstract}



\section{Introduction}
\label{intro}

Active galactic nuclei (AGN) are the most powerful non-transient objects in the Universe \citep{OF06}. Their luminosity ranges from $10^{40}$ to $10^{47}$~erg\,s$^{-1}$. 
Blazars form a sub-class of AGN, containing optically violently variable (OVV) quasars, high-polarization quasars (HPQ) and BL Lacertae (BL Lac) objects. The latter are characterized by weaker or no emission lines and extreme variability \citep{Ant93,Urry,Zen97}.
Blazars are all radio-loud AGN and often show apparent superluminal motion in their jet components \citep[see the recent review by e.g.][]{Per13}. 
The observed luminosity of blazars is significantly magnified by Doppler boosting, an effect of the relativistically beamed emission from the parsec-scale jet components \citep{BK79,KP81,OB82,Lob98,Kel04}. This prominent character makes the blazars excellent laboratories to study AGN phenomena.

The tremendous energy release process of blazars is known to be associated with the accretion of material into a central supermassive black hole (SMBH) at the heart of the galaxy. It happens in an extremely compact region surrounding the SMBH.
The radio properties of blazars are mostly manifested by highly relativistic jets launched from the vicinity of the central power engine \citep{Rees71,Sch74}. High-resolution radio observations of jetted outflows using the technique of Very Long Baseline Interferometry \citep[VLBI,][]{Whi71} shed light on the immediate environments of SMBHs, provide direct measurements of kinematic and magnetic properties of the jet flow \citep[e.g.][]{Hada11,Doe12}, and allow testing jet models \citep[e.g.][]{MG85,Fer88,Mar09,PC12}.

Although blazars only account for a small fraction in the general AGN population, owing to their high observed luminosity, they often dominate the flux-density-limited AGN survey samples.
VLBI surveys of radio-loud core-dominated AGN provide fundamental information on the statistical properties of the compact AGN jets and also form the basis for further detailed observations of individual objects. Several large VLBI surveys at various frequencies have been carried out in the past three decades or so, resulting in an imaging data base of several thousand AGN (for examples, see the references listed in Table~\ref{tab1}). Ground-based VLBI at mm wavelengths provides a typical angular resolution of $\lesssim0.2$~mas (at 7~mm), thus offering the best tool for imaging the fine radio structures of compact AGN jets on sub-parsec scales, approaching the site where the jet is launched and collimated.

The highest-resolution VLBI imaging  currently available is at 86 GHz \citep[e.g.,][]{Lee08}, although this may change in the fairly near future as 230-GHz observations with better baseline coverage become available \citep[][]{Kri13}. The first single-baseline interference fringes demonstrating the feasibility of 3-mm VLBI were detected at 89~GHz \citep{Rea83}. Since then, a number of VLBI observations of individual sources and different samples have been conducted at 86~GHz \citep[][and references therein]{Bea97,Ran98,Lon98,Lob00,Lee08}.
These observations have led to the detection of 126 sources in total, of which 114 have high ($\sim$10 microarcsecond-level) resolution images \citep{Ran98,Lob00,Lee08}. In fact, these sources constitute a group of the {\it brightest and most compact} radio AGN.

Apart from increasing the observing frequency, the angular resolution can also be improved by expanding the maximum baseline length of the VLBI network. This basic idea led to the development of space VLBI (SVLBI) which involves one or more radio telescopes installed on Earth-orbiting spacecraft. The ground-based VLBI stations and the space radio telescope(s) together form a space--ground VLBI network \citep{Bur84}. Currently, ground-only VLBI does not have sufficient resolving power to study the regions closest to the SMBHs where the jets are created and collimated. The extension of VLBI baselines into space offers the unique opportunity to probe these central regions of AGN by direct imaging.

In the 1980s, the first SVLBI experiments were successfully performed with a geostationary Tracking and Data Relay Satellite (TDRS) and some ground radio telescopes at 2.3 and 15~GHz \citep{Levy86,Lin89,Lin90}. The first dedicated SVLBI mission was the VLBI Space Observatory Programme (VSOP) led by Japan \citep{Hir00a}. The {\it HALCA} satellite carrying an 8-m diameter radio telescope on board was launched in 1997 \citep{Hir98}. The space radio telescope worked at 1.6 and 5~GHz. The apogee altitude of HALCA was  21\,400 km, providing an angular resolution up to three times higher than available with the ground-only VLBI at the same observing frequency. The VSOP conducted an extensive 5-GHz AGN survey targeting a sample of 402 sources above 1~Jy flux density \citep{Hir00b}. The results indicate that the AGN cores have an average diameter of 0.2~mas, and 14 per cent of the cores are smaller than 0.04~mas \citep{Hor04,Sco04}.

In order to explore the innermost jets of the extremely compact AGN, SVLBI at higher frequency and with increased apogee height is required. 
Initial studies of a second-generation Japanese SVLBI mission VSOP-2 including observations at 8, 22 and 43 GHz and providing a maximum resolution of 38 microarcseconds \citep[38\,$\mu$as;][]{Hir00c,Tsu09} were carried out, but it was decided not to take the project forward due to technical difficulties in meeting some of the required specifications.
Currently the {\it RadioAstron} mission led by Russia is operational \citep{Kar13}. The satellite, launched with a 10-m orbiting radio telescope in 2011, provides baselines up to $\sim$300\,000~km to the ground radio telescopes. This in principle allows a resolution of $\sim$7~$\mu$as at the highest of the four observing frequencies (0.3, 1.6, 5 and 22~GHz). Such a fine resolution could be obtained when the spacecraft is at around its apogee. 
However, the imaging capability of RadioAstron is rather limited because a large fraction of the orbital period is spent at around the apogee due to the very eccentic elliptical orbit.

Future SVLBI missions would leap forward in the direction of improving both the resolution and the imaging capability. 
For instance, Russia is planning a millimeter-wavelength space-VLBI project named {\it Millimetron}, including a 12-m space radio telescope working at the highest frequency of 4700~GHz \citep{Kar07,Wild}. 
Recently, the Shanghai Astronomical Observatory (SHAO) of the Chinese Academy of Sciences (CAS) started studying such a program \citep{Hong}. The objective is to achieve ultrahigh angular resolution (a few $\mu$as) to image the immediate vicinity of the SMBHs, to study the black hole physics and to test the general relativity in strong gravity field. The first phase of the program involves two satellites, each carrying a 10-m radio telescope, operating at the highest frequency of 43~GHz. The proposed apogee height of the satellites is $\sim$60\,000~km. The two space telescopes, working together with ground-based VLBI stations, could provide angular resolution as fine as 20~$\mu$as, together with a favourable $(u,v)$ coverage for imaging.

To fulfil the scientific objectives of the SVLBI array, it is essential to have a sample of AGN that are sufficiently compact and bright for successful detection and imaging. Assembling a fundamental target list is vital for several reasons: calibration of the space telescope (pointing, sensitivity), investigations of jet physics (formation and acceleration of the relativisitic jets, magnetic fields in inner jets, transverse structure of jet flow, etc.), and studies of the statistical properties of the most compact AGN cores. The goal of this paper is to supplement the AGN lists known from previous 86-GHz ground-based VLBI surveys with potential new compact targets that have never been imaged at this frequency. The frequency of 86~GHz was chosen because {\it (i)} this is currently the highest where VLBI imaging data on a sufficiently large sample of sources are available, and {\it (ii)} the angular resolution of the ground-only VLBI networks is almost as high as the resolution of the proposed space--ground array at 43~GHz. A sample of compact and bright radio sources, to be confirmed with ground-based 86-GHz VLBI imaging, offers a good basis for a sample to be targeted later with 43-GHz SVLBI imaging. 

With the recent release of high-frequency all-sky AGN flux density data bases, such as the point source catalogues of the {\it Wilkinson Microwave Anisotropy Probe} \citep[{\it WMAP},][]{Ben03} and the {\it Planck} \citep{Pla11a} space telescopes, it became possible to search for more candidates. The single-dish flux densities measured at several high-frequency bands, together with public VLBI imaging data at lower frequencies, allow us to identify new bright and compact AGN suitable as targets for the future mm-wavelength SVLBI. On the other hand, ground-based VLBI imaging of these sources at 86~GHz would significantly extend the samples available at present. In Section~\ref{sample}, we describe the sample selection method and give the result. Section~\ref{statistics} presents some statistical properties of our new sample. Section \ref{summary} gives a summary of the findings and outlines our plans for continuing this work.

\section{The sample selection}
\label{sample}

\begin{table}
\caption{Major VLBI imaging surveys of radio-loud AGN, with the 86-GHz surveys at the bottom.}
\begin{center}
\begin{tabular}{cccc}
\hline\hline
Survey       & $\nu$ (GHz) &  $N$ & Ref. \\
\hline
PR           & 5        & 37       & 1\\
CJF          & 1.6, 5   & 293      & 2--7 \\
VSOP PLS     & 5        & 374      & 8 \\
VSOP         & 5        & 242      & 9--11 \\
VCS          & 2.3, 8.4 & 5756     & 12--17 \\
RRFID        & 2.3, 8.4 & 782 & 18--20 \\
VIPS         & 5        & 1127     & 21 \\
VLBA 2cm / MOJAVE & 15  & 259      & 22--27 \\
CRF          & 24, 43   & 274      & 28 \\  
mJIVE-20     & 1.4      & $>$4300  & 29 \\
\hline
Rantakyr\"o+ & 86       & 13       & 30 \\
Lobanov+     & 86       & 17       & 31 \\
Lee+         & 86       & 109      & 32 \\
\hline
\end{tabular}
\end{center}
\label{tab1}
Notes. Col.~1: abbreviated survey name, Col.~2: observing frequency in GHz, Col.~4: number of sources imaged, Col.~4: references   
1. \citet{PR88};
2--7. \citet{Pol95,Tha95,Xu95,Tay94,Hen95,Pol03};
8. \citet{Fom00};
9--11. \citet{Hir00b,Sco04,Dod08};
12--17. \citet{Bea02,Fom03,Pet05,Pet06,Kov07,Pet08};
18--20. \citet{Fey96,Fey97,Fey00};
21. \citet{Hel07}
22--27. \citet{Kel98,Zen02,Kov05,Lis05,Lis09,Lis13};
28. \citet{Cha10};
29. \citet{DM14};
30. \citet{Ran98};
31. \citet{Lob00};
32. \citet{Lee08}. 
\end{table}

An important practical question about the future SVLBI, in particular the proposed Chinese mm-SVLBI array, is how many targets can be detected above 100 mJy ($\sim 7\sigma$)  at the highest frequency band (43~GHz) on the longest achievable space--ground baselines of $\sim$70\,000 km, assuming state-of-the-art receivers and a maximum bandwidth of 256~MHz \citep{Kel13}. An analysis of the results of earlier VLBI surveys, and eventually a dedicated pre-launch AGN survey using ground-based VLBI arrays at the highest possible frequencies should yield a realistic estimate to answer this question.

Undoubtedly, the pre-launch AGN survey sample should contain the brightest and most compact sources known. 
Several large VLBI surveys of AGN were carried out in the past, working at cm and mm wavelengths \citep[see the reviews in e.g.][]{Fre06,Kov09}. All of them were limited to a certain flux density, thus preferentially targeting the bright AGN. The improved sensitivity of the instruments with time led to lower flux density limits and the increase in the sample sizes. Table~\ref{tab1} lists some representative AGN imaging surveys and gives references for their details. The 86-GHz surveys are listed at the end of the table.  
 
The number of VLBI-imaged sources at mm wavelengths is small compared to the sample sizes of cm surveys. The moderate size of the presently known 86-GHz AGN sample can partly be attributed to technical difficulties: the small number and the limited availability of VLBI telescopes operating at 3~mm, the relatively poor baseline sensitivity, and the short atmospheric coherence time at 86~GHz. On the other hand, the continuum radio spectra steepening at high frequencies contribute to the decrease of suitably strong targets. 

So far, VLBI surveys at 86~GHz resulted in images of just over 100 bright and most compact AGN. A first step towards an enlarged sample suitable for observations with a future mm-SVLBI array is to find further candidates 
which can then be confirmed as compact radio AGN using ground mm-VLBI observations. 
To this end, we turned to the {\it WMAP} and {\it Planck} point source catalogues. 

The {\it WMAP} performed its all-sky survey at five different frequency bands from 23 to 94~GHz, from 2001 August to 2010 August. The primary goal of the mission was to study the properties of the Cosmic Microwave Background (CMB) radiation. To remove the foreground radio emission, a catalogue of discrete sources had to be created \citep[e.g.][]{Chen09}. The {\it Planck} spacecraft complemented the CMB survey with improved angular resolution, sensitivity and frequency coverage, from 2009 August to 2013 October. The {\it Planck} covered a broader frequency range, in 9 different bands from 30 to 857 GHz. The range of the upper four frequency bands of the {\it WMAP} overlapped with the lowest four bands of the {\it Planck}. 

Recently, \citet{Chen13} compared the {\it WMAP} point source data from the first 7 years and the early release of the {\it Planck} compact source catalogue \citep{Pla11b}. They compiled a list of 198 extragalactic sources detected with both space telescopes. The identification was based on the matching positions, no flux density limit was set, and the distribution of the sources is uniform over the whole sky (the sources within 5\degr of the Galactic plane were excluded to avoid confusion). Thus the \citet{Chen13} list can be regarded as an unbiased sample of bright extragalactic radio sources at mm wavelengths. The 198 sources include 161 blazars (135 flat-spectrum radio quasars and 26 BL Lacertae objects), 28 non-blazar AGN, 2 galaxies, and 7 unidentified radio sources \citep{Chen13}.
  
To define a new sample of bright and compact AGN potentially suitable for mm-wavelength SVLBI, we applied the following four selection criteria: 
\begin{description}
\item[(a)] A flat radio continuum spectrum (flux density $S_\nu \propto \nu^{\alpha}$, with a spectral index $\alpha \geq -0.5$) between 33 and 94~GHz.
\item[(b)] The object is classified as either a flat-spectrum radio quasar or a BL Lac object \citep{Chen13}. Other types of sources such as starburst galaxies, normal galaxies, extended radio galaxies, and Galactic objects are ruled out this way.
\item[(c)] The declination is at least $-40\degr$, for the sake of good imaging together with the ground-based VLBI telescopes mostly located in the northern hemisphere.
\item[(d)] The source has never been imaged with ground-based VLBI at 86~GHz.
\end{description}
 
The flatness of the radio spectrum used for criterion (a) is believed to result from the superposition of several compact sub-components in the AGN ``core'' (i.e. the base of the compact jet) which are synchrotron self-absorbed with different turnover frequencies. Therefore the flat overall spectrum is a good indication of the compactness of the radio structure. The selection criteria (c) and (d) are the same as used for previous studies \citep{Lob00,Lee08}. Here we do not directly apply a lower flux density threshold, as was done e.g. by \citet{Lee08} who constrained their sample to $S_{86} > 0.3$~Jy. Note however, that the combined {\it WMAP--Planck} sample of \citet{Chen13} does have an implicit lower threshold of the flux density (somewhat below 1~Jy) because of the point-source detection limits of the space telescopes.

After applying the four filters above, 49 sources remain from the total of 198 {\it WMAP--Planck} objects \citep{Chen13}. 
Table~\ref{tab2} lists the basic information of these 49 sources: the source names derived from the J2000 and B1950 coordinates, respectively, right ascension ($\alpha_{\rm J2000}$), declination ($\delta_{\rm J2000}$), spectroscopic redshift ($z$), and type (BLL means BL Lac, FSRQ abbreviates flat-spectrum radio quasar). The coordinates and the redshifts are taken from the NASA/IPAC Extragalactic Database\footnote{http://ned.ipac.caltech.edu/} (NED). The object types are adopted from \citet{Chen13}.

In their recent work, \citet{GF11} selected 37 compact AGN as potential SVLBI calibrators from the five-year WMAP point source catalogue of \citet{Chen09}. They used three selection criteria: a total flux density limit of $S_{86}>1$~Jy, declination above $-40\degr$, and no previous 86-GHz VLBI observation of the sources. The last two criteria are the same as we applied in (c) and (d), respectively. Among the 37 sources of \citet{GF11}, 14 overlap with our sample. The differences in the sample size and content could arise from the different parent catalogues, and the other selection criteria: the flux density limit used by \citet{GF11} on the one hand, and the flat radio spectrum required in our paper on the other.


\section{Properties of the sample}
\label{statistics}

\begin{figure}
\includegraphics[scale=0.45]{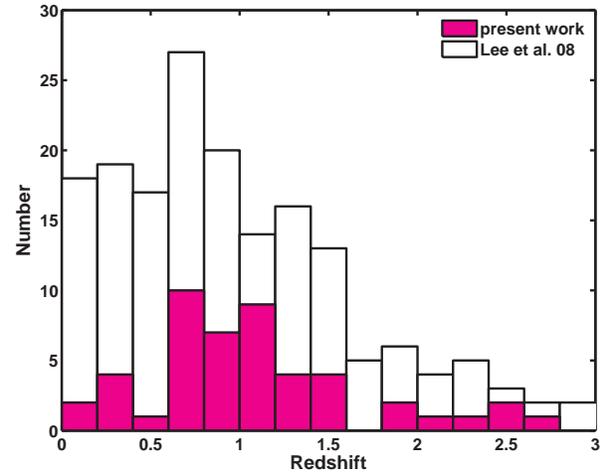}
\caption{Histogram of the redshift distribution for the 86-GHz global VLBI imaging survey \citep[][open bars]{Lee08} combined with our present sample of 48 candidate compact sources (filled bars).}
\label{fig:redshift}
\end{figure}

Except for J0629$-$1959, all the sources in the sample are prominent optical AGN with measured spectroscopic redshift available in the literature (Table~\ref{tab2}). The lowest redshift is 0.049 (for J1517$-$2422, also known as AP~Lib), the highest value is 2.680 (for J0242+1101), the median redshift is 1.0. The histogram of the redshift distribution is shown in Fig.~\ref{fig:redshift} shown as filled bars. The data of the 86-GHz global VLBI imaging survey \citep{Lee08} are also displayed with open bars. The comparison indicates that the redshift distribution of the new sources is in good agreement with those already shown to be suitable for 86-GHz VLBI observations.
There is a generally decreasing trend with increasing $z$, as a natural consequence of the flux-density-limited samples. From the largest cosmological distances, only the few most luminous sources enter in the sample. At high redshifts, which also correspond to the era preceding the period of the peak AGN activity in the Universe (at $z$ between $\sim$0.5 and 2), the number of sources gradually declines.

Our sample contains 7 BL Lacs (14 per cent) and 42 flat-spectrum radio quasars (86 per cent). The percentage of the BL Lacs is consistent with that in the parent {\it WMAP--Planck} sample \citep[16 per cent,][]{Chen13}. It is comparable to the 19 per cent found in the 86-GHz VLBI sample of \citet{Lee08}. Interestingly, another large VLBI imaging survey, the MOJAVE 15-GHz sample \citep{Lis13} with a total of 183 radio-loud AGN also contains 16 per cent BL Lacs. The ratio of BL Lac objects in the radio-selected blazar samples is an important input for understanding the difference between BL Lacs and FSRQs \citep[e.g.][]{Gio12}. 
Based on the {\it Planck} flux densities interpolated to the four {\it WMAP} frequency bands by \citet{Chen13}, we calculated the spectral index $\alpha$ of each source by fitting a power-law continuum 
using the measured flux densities at 33, 41, 61 and 94 GHz. 
This parameter was then used to select the flat-spectrum sources with $\alpha \geq -0.5$ as described in Section~\ref{sample}. 

\begin{figure}
\includegraphics[width=0.23\textwidth]{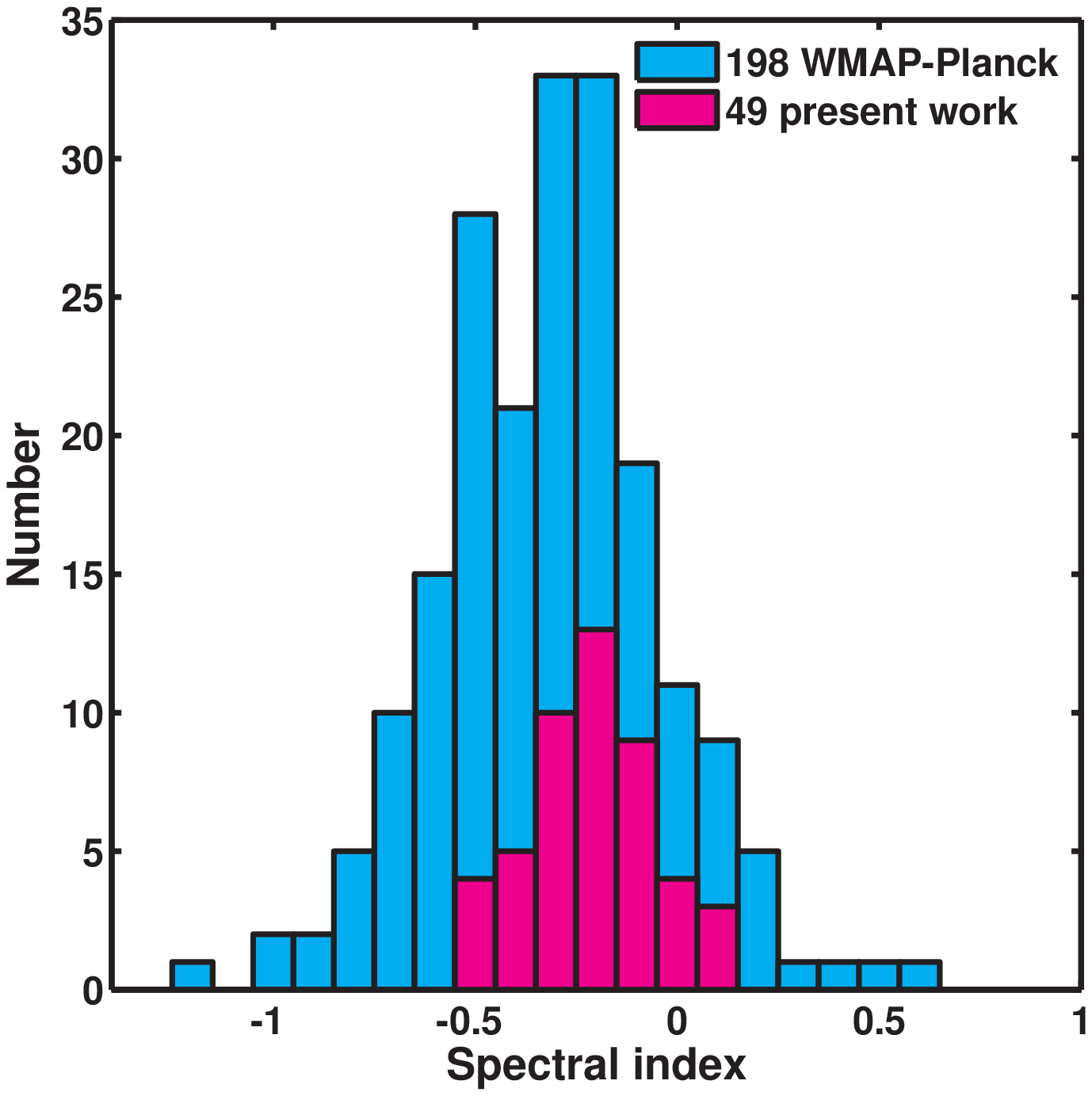}
\includegraphics[width=0.23\textwidth]{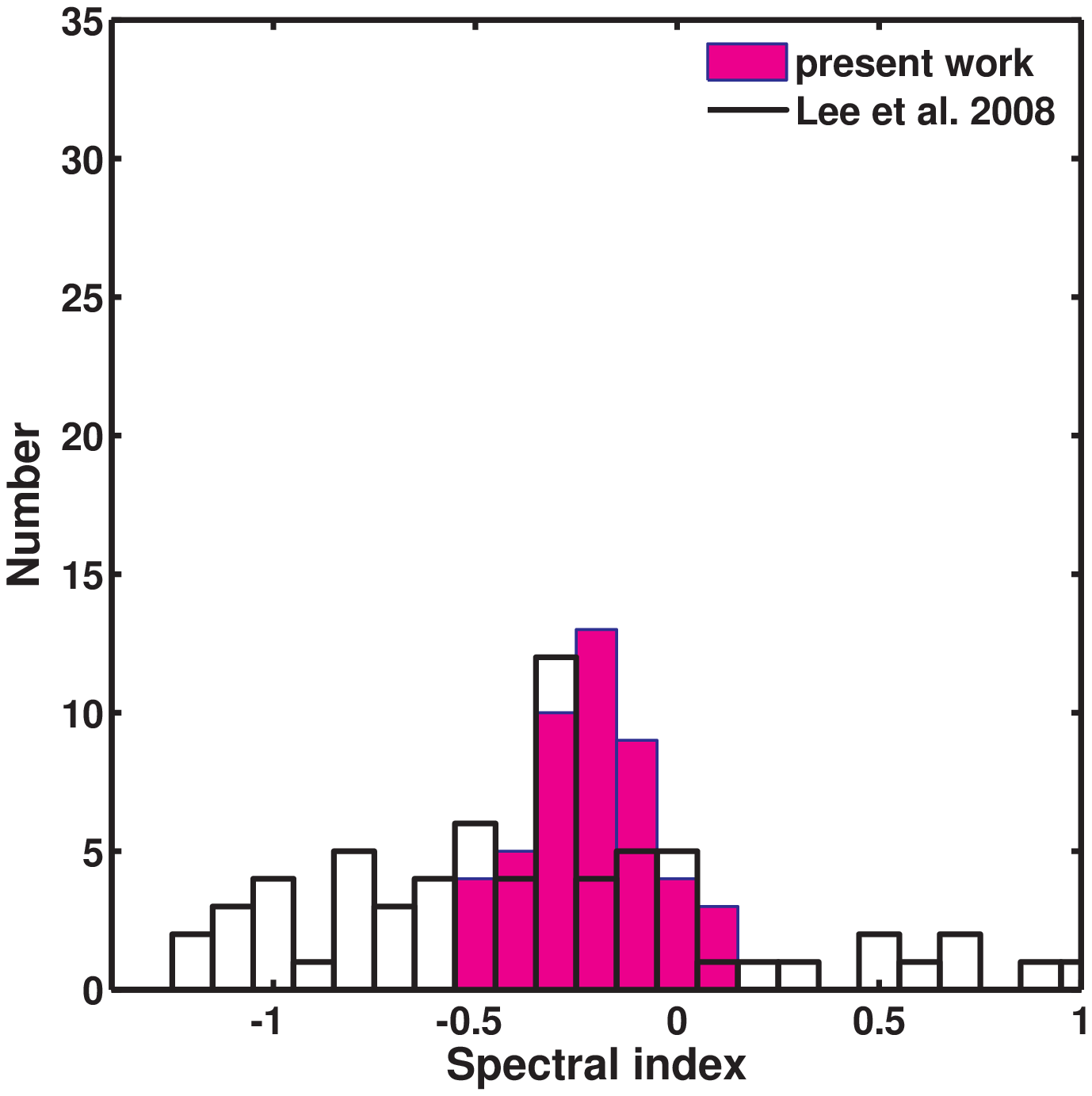}
\caption{Plots of the spectral index distribution in our sample of 49 blazars (red) overlapped with the parent {\it WMAP--Planck} sample \citep[][blue, left panel]{Chen13}. The spectral index distribution for the \citet{Lee08} 86-GHz VLBI sample is also shown as open bars in the right panel, with the data adopted from \citet{GF11}.  
}
\label{fig:sindex}
\end{figure}

Figure~\ref{fig:sindex} shows the distribution of the spectral index in the present sample, compared with others. The maximum spectral index in our sample is $\alpha=0.13$, the corresponding radio source (J0719$+$3307) shows a rising spectrum from cm to mm wavelengths. The median spectral index in the sample is $-0.22$.
The spectral indices in the \citet{Lee08} sample (right panel of Fig.~\ref{fig:sindex}) show a broader distribution, while its peak is consistent with that of the present sample, although we filtered out the steep-spectrum sources with selection criterion (a). The objects with steep spectra typically correspond to the radio galaxies in \citet{Lee08}.

\begin{figure}
\includegraphics[width=0.4\textwidth]{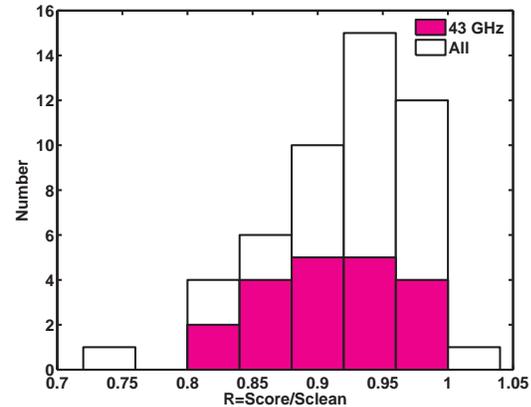}
\caption{Histogram of the source compactness parameter $R = S_{\rm core}/S_{\rm CLEAN}$ in our sample. The measurements at 43~GHz are marked with red colour. 
}
\label{fig:sratio}
\end{figure}

Using publicly available\footnote{http://astrogeo.org} calibrated visibility data from earlier VLBI surveys (Table~\ref{tab1}), we investigated the compactness of the 49 sources listed in Table~\ref{tab2}. We used the highest-frequency data available for a given object, at 8, 15, 24 or 43~GHz. Among the possible methods to assess the compactness, we adopted the one applied by \citet{Lee08}, by calculating the ratio of the core flux density to the total flux density of the CLEAN components obtained during the imaging, $R= S_{\rm core}/S_{\rm CLEAN}$. We obtained $S_{\rm core}$ by fitting the brightest core component with a circular Gaussian brightness distribution model in {\sc Difmap} \citep{She97}, and $S_{\rm CLEAN}$ by integrating the flux density contained in the visible emission region in the VLBI image. Since the source compactness obviously depends on the observing frequency \citep[e.g.][]{Cha10}, and the structures tend to be smaller at higher frequencies, the heterogeneous $R$ values obtained at different frequencies are not directly comparable in our sample. However,  for most of the sources, the core dominance was at least 90 per cent (Fig. \ref{fig:sratio}), indicating that the members of the present sample are indeed compact radio-loud AGN. The minimum and median values for $R$ were 0.76 and 0.95, respectively. By visual inspection, the overall core dominance of the objects in the sample is consistent with the appearance of the sources in the lower-frequency VLBI images, where they all show compact core--jet structures or naked cores, typical for  
radio-loud core-dominated AGN. 

\begin{table*}
\caption{The catalogue of 49 new candidate compact blazars for mm-SVLBI observations.}
\begin{tabular}{cccccccc}
\hline 
No. & Name (J2000) & Name (B1950) & $\alpha_{\rm J2000}$ & $\delta_{\rm J2000}$ & $z$  & $\alpha$ & Type \\ 
\hline 
1  &J0137$-$2430	&0135$-$247 &01 37 38.3465	&$-$24 30 53.886	&0.838	&$-0.06$ &FSRQ\\   
2  &J0242$+$1101	&0239$+$108 &02 42 29.1709	&$+$11 01 00.728	&2.680	&$-0.25$ &FSRQ\\   
3  &J0406$-$3826	&0405$-$385	&04 06 59.0353	&$-$38 26 28.042	&1.285	&$-0.23$ &FSRQ\\   
4  &J0428$-$3756	&0426$-$380	&04 28 40.4243	&$-$37 56 19.581	&1.110	&$-0.22$ &BLL \\   
5  &J0449$+$1121	&0446$+$112	&04 49 07.6711	&$+$11 21 28.596	&1.207	&$-0.14$ &FSRQ\\   
6  &J0453$-$2807	&0451$-$282	&04 53 14.6468	&$-$28 07 37.327	&2.559	&$-0.18$ &FSRQ\\   
7  &J0457$-$2324	&0454$-$234	&04 57 03.1792	&$-$23 24 52.020	&1.003	&$-0.18$ &FSRQ\\   
8  &J0607$-$0834	&0605$-$085	&06 07 59.6992	&$-$08 34 49.978	&0.872	&$-0.23$ &FSRQ\\   
9  &J0629$-$1959	&0627$-$199	&06 29 23.7619	&$-$19 59 19.724	&... 	  &$-0.29$ &BLL \\   
10 &J0634$-$2335	&0632$-$235	&06 34 59.0010	&$-$23 35 11.957	&1.535	&$-0.33$ &FSRQ\\   
11 &J0719$+$3307	&0716$+$332	&07 19 19.4197	&$+$33 07 09.709	&0.779	&$ 0.13$ &FSRQ\\   
12 &J0725$-$0054	&0723$-$008	&07 25 50.6400	&$-$00 54 56.544	&0.128	&$ 0.01$ &BLL \\   
13 &J0757$+$0956	&0754$+$100	&07 57 06.6429	&$+$09 56 34.852	&0.266	&$ 0.01$ &BLL \\   
14 &J0808$-$0751	&0805$-$077	&08 08 15.5360	&$-$07 51 09.887	&1.837	&$-0.16$ &FSRQ\\   
15 &J0824$+$3916	&0821$+$394	&08 24 55.4839	&$+$39 16 41.904	&1.216	&$-0.22$ &FSRQ\\   
16 &J0920$+$4441	&0917$+$449	&09 20 58.4585	&$+$44 41 53.985	&2.186	&$-0.32$ &FSRQ\\   
17 &J0921$-$2618	&0919$-$260	&09 21 29.3539	&$-$26 18 43.386	&2.300	&$-0.28$ &FSRQ\\   
18 &J0927$+$3902	&0923$+$392	&09 27 03.0139	&$+$39 02 20.852	&0.695	&$-0.47$ &FSRQ\\   
19 &J0927$-$2034	&0925$-$203	&09 27 51.8243	&$-$20 34 51.233	&0.347	&$ 0.05$ &FSRQ\\   
20 &J1033$+$4116	&1030$+$415	&10 33 03.7079	&$+$41 16 06.233	&1.118	&$-0.39$ &FSRQ\\   
21 &J1037$-$2934	&1034$-$293	&10 37 16.0797	&$-$29 34 02.813	&0.312	&$-0.03$ &FSRQ\\   
22 &J1056$+$7011	&1053$+$704	&10 56 53.6175	&$+$70 11 45.916	&2.492	&$ 0.06$ &FSRQ\\   
23 &J1058$+$0133	&1055$+$018	&10 58 29.6052	&$+$01 33 58.824	&0.890	&$-0.30$ &FSRQ\\   
24 &J1127$-$1857	&1124$-$186	&11 27 04.3924	&$-$18 57 17.442	&1.050	&$-0.41$ &FSRQ\\   
25 &J1130$-$1449	&1127$-$145	&11 30 07.0526	&$-$14 49 27.388	&1.184	&$-0.12$ &FSRQ\\   
26 &J1146$+$3958	&1144$+$402	&11 46 58.2979	&$+$39 58 34.304	&1.090	&$-0.38$ &FSRQ\\   
27 &J1337$-$1257	&1334$-$127	&13 37 39.7828	&$-$12 57 24.693	&0.539	&$-0.23$ &FSRQ\\   
28 &J1357$+$1919	&1354$+$195	&13 57 04.4367	&$+$19 19 07.372	&0.720	&$-0.28$ &FSRQ\\   
29 &J1517$-$2422	&1514$-$241 &15 17 41.8131	&$-$24 22 19.476	&0.049	&$-0.17$ &BLL \\   
30 &J1626$-$2951	&1622$-$297	&16 26 06.0208	&$-$29 51 26.971	&0.815	&$-0.07$ &FSRQ\\   
31 &J1637$+$4717	&1636$+$473	&16 37 45.1306	&$+$47 17 33.831	&0.740	&$-0.31$ &FSRQ\\   
32 &J1727$+$4530	&1726$+$455	&17 27 27.6508	&$+$45 30 39.731	&0.717	&$-0.32$ &FSRQ\\   
33 &J1734$+$3857	&1732$+$389	&17 34 20.5785	&$+$38 57 51.443	&0.970	&$-0.13$ &FSRQ\\   
34 &J1751$+$0939	&1749$+$096	&17 51 32.8186	&$+$09 39 00.728	&0.322	&$-0.21$ &BLL \\   
35 &J1801$+$4404	&1800$+$440	&18 01 32.3148	&$+$44 04 21.900	&0.663	&$-0.50$ &FSRQ\\   
36 &J1849$+$6705	&1849$+$670	&18 49 16.0723	&$+$67 05 41.680	&0.657	&$-0.17$ &FSRQ\\   
37 &J1911$-$2006	&1908$-$201	&19 11 09.6529	&$-$20 06 55.109	&1.119	&$-0.12$ &FSRQ\\   
38 &J1923$-$2104	&1920$-$211	&19 23 32.1898	&$-$21 04 33.333	&0.874	&$-0.06$ &FSRQ\\   
39 &J1957$-$3845	&1954$-$388	&19 57 59.8193	&$-$38 45 06.356	&0.630	&$-0.50$ &FSRQ\\   
40 &J2000$-$1748	&1958$-$179	&20 00 57.0904	&$-$17 48 57.673	&0.650	&$-0.30$ &FSRQ\\   
41 &J2006$+$6424	&2005$+$642	&20 06 17.6946	&$+$64 24 45.418	&1.574	&$-0.44$ &FSRQ\\   
42 &J2031$+$1219	&2029$+$121	&20 31 54.9943	&$+$12 19 41.340	&1.215	&$-0.33$ &BLL \\   
43 &J2101$-$2933	&2058$-$297	&21 01 01.6600	&$-$29 33 27.836	&1.492	&$-0.06$ &FSRQ\\   
44 &J2148$+$0657	&2145$+$067	&21 48 05.4587	&$+$06 57 38.604	&0.990	&$-0.45$ &FSRQ\\   
45 &J2203$+$1725	&2201$+$171	&22 03 26.8937	&$+$17 25 48.248	&1.075	&$-0.08$ &FSRQ\\   
46 &J2229$-$0832	&2227$-$088	&22 29 40.0843	&$-$08 32 54.436	&1.560	&$-0.04$ &FSRQ\\   
47 &J2232$+$1143	&2230$+$114	&22 32 36.4089	&$+$11 43 50.904	&1.037	&$-0.35$ &FSRQ\\   
48 &J2246$-$1206	&2243$-$123	&22 46 18.2320	&$-$12 06 51.278	&0.632	&$-0.23$ &FSRQ\\   
49 &J2311$+$3425	&2308$+$341	&23 11 05.3288	&$+$34 25 10.906	&1.817	&$-0.17$ &FSRQ\\    
\hline
\end{tabular}
\label{tab2}
\end{table*}

\section{Summary and outlook}
\label{summary}

We selected a complementary sample of 49 bright and compact radio AGN as potential targets for future mm VLBI and SVLBI observations. The sources were drawn form the common {\it WMAP--Planck} source list compiled by \citet{Chen13}, having flat radio spectrum and declination above $-40\degr$. The latter criterion ensures their visibility from northern-hemisphere VLBI arrays. The new candidates have never been imaged with ground-based VLBI at the highest observing frequency used for large AGN surveys (86~GHz). Based on their optical identification, continuum spectral properties, and compactness in lower-frequency VLBI surveys, these sources appear promising candidates for successful VLBI detection with ground-based VLBI at 86~GHz or SVLBI at 43~GHz in the future. This study is a part of the efforts to define a target sample for a  proposed Chinese SVLBI mission involving two spaceborne radio telescopes, operating at frequencies up to 43~GHz \citep{Hong}. 

The sample presented in this paper would increase the number of extragalactic objects covered in past 86-GHz VLBI surveys \citep{Ran98,Lob00,Lee08} by nearly 40 per cent. The next immediate step is to perform VLBI fringe detection tests with these objects, and eventually long-baseline imaging observations  to confirm their compactness. In the near future, further improvement in selecting new candidates is expected from the use of the final release of the {\it Planck} catalogue of compact sources which comprises more sources and more accurate flux density measurements. The selection criteria applied in the present paper could be adopted to find additional compact AGN candidates with somewhat lower flux densities.

\section*{Acknowledgements}
The authors are grateful for the suggestions of the referee which significantly improved the paper.
We thank the constructive comments given by the members of the AGN working group of the Chinese Space VLBI Array program.
This work was supported by the Strategic Priority Research Program on Space Science of the Chinese Academy of Sciences (CAS; grant No. XDA04060700), the 973 Program (grant No. 2013CB837900), NSFC (grant No. 11261140641, U1331205), CAS (grant No. KJZD-EW-T01), and the China--Hungary Collaboration and Exchange Programme by the International Cooperation Bureau of the CAS.
SF acknowledges the support from the Hungarian Scientific Research Fund (OTKA K104539 and NN110333).
This research has made use of the NASA/IPAC Extragalactic Database (NED) which is operated by the Jet Propulsion Laboratory, California Institute of Technology, under contract with the National Aeronautics and Space Administration.

\label{lastpage}

\end{document}